\documentclass[floatfix,10pt,twocolumn,superscriptaddress,aps,prd,nofootinbib]{revtex4-2}
\usepackage{graphicx}
\usepackage{subfigure}
\usepackage{dcolumn}\usepackage{bm}
\usepackage[usenames]{color}
\usepackage{lineno}
\usepackage{amsmath}
\usepackage{amssymb}
\usepackage{etoolbox} 
\usepackage{url} 
\usepackage{placeins}
\usepackage{hyperref}

\usepackage{multirow} 
\usepackage{orcidlink}

\newcommand*\linenomathpatch[1]{%
  \cspreto{#1}{\linenomath}%
  \cspreto{#1*}{\linenomath}%
  \csappto{end#1}{\endlinenomath}%
  \csappto{end#1*}{\endlinenomath}%
}
\linenomathpatch{equation}
\linenomathpatch{gather}
\linenomathpatch{multline}
\linenomathpatch{align}
\linenomathpatch{alignat}
\linenomathpatch{flalign}

\def\address{\@ifstar{\address@star}%
  {\@ifnextchar[{\address@optarg}{\address@noptarg}}}

\newcommand\ee{e^+e^-}

\newcommand\ainv{A'\to invisible}

\newcommand\g{\gamma}
\newcommand\ma{m_{A'}}
\newcommand\mc{m_\chi}

\newcommand\Na{{N}_{A'}}

\begin{document}

\title{EUROPEAN LABORATORY FOR PARTICLE PHYSICS\\
\vskip0.5cm
\hspace{-3.2cm}{\rightline{\rm  CERN-EP-2023-130}}
\vskip1.0cm
Search for Light Dark Matter  with NA64 at CERN}

\author{Yu.~M.~Andreev\orcidlink{0000-0002-7397-9665}}
\affiliation{Authors affiliated with an institute covered by a cooperation agreement with CERN}
\author{D.~Banerjee\orcidlink{0000-0003-0531-1679}}
\affiliation{CERN, European Organization for Nuclear Research, CH-1211 Geneva, Switzerland}
\author{B.~Banto Oberhauser\orcidlink{0009-0006-4795-1008}}
\affiliation{ETH Z\"urich, Institute for Particle Physics and Astrophysics, CH-8093 Z\"urich, Switzerland}
\author{J.~Bernhard\orcidlink{0000-0001-9256-971X}}
\affiliation{CERN, European Organization for Nuclear Research, CH-1211 Geneva, Switzerland}
\author{P.~Bisio\orcidlink{/0009-0006-8677-7495}}
\affiliation{INFN, Sezione di Genova, 16147 Genova, Italia}
\affiliation{Universit\`a degli Studi di Genova, 16126 Genova, Italia}
\author{A.~Celentano\orcidlink{0000-0002-7104-2983}}
\affiliation{INFN, Sezione di Genova, 16147 Genova, Italia}
\author{N.~Charitonidis\orcidlink{0000-0001-9506-1022}}
\affiliation{CERN, European Organization for Nuclear Research, CH-1211 Geneva, Switzerland}
\author{A.~G.~Chumakov\orcidlink{/0000-0002-6012-2435}}
\affiliation{Authors affiliated with an institute covered by a cooperation agreement with CERN}
\author{D.~Cooke}
\affiliation{UCL Departement of Physics and Astronomy, University College London, Gower St. London WC1E 6BT, United Kingdom}
\author{P.~Crivelli\orcidlink{0000-0001-5430-9394}}
\affiliation{ETH Z\"urich, Institute for Particle Physics and Astrophysics, CH-8093 Z\"urich, Switzerland}
\author{E.~Depero\orcidlink{0000-0003-2239-1746}}
\affiliation{ETH Z\"urich, Institute for Particle Physics and Astrophysics, CH-8093 Z\"urich, Switzerland}
\author{A.~V.~Dermenev\orcidlink{0000-0001-5619-376X}}
\affiliation{Authors affiliated with an institute covered by a cooperation agreement with CERN}
\author{S.~V.~Donskov\orcidlink{0000-0002-3988-7687}}
\affiliation{Authors affiliated with an institute covered by a cooperation agreement with CERN}
\author{R.~R.~Dusaev\orcidlink{0000-0002-6147-8038}}
\affiliation{Authors affiliated with an institute covered by a cooperation agreement with CERN}
\author{T.~Enik\orcidlink{0000-0002-2761-9730}}
\affiliation{Authors affiliated with an international laboratory covered by a cooperation agreement with CERN}
\author{V.~N.~Frolov}
\affiliation{Authors affiliated with an international laboratory covered by a cooperation agreement with CERN}
\author{R.~B.~Galleguillos~Silva}
\affiliation{Center for Theoretical and Experimental Particle Physics, Facultad de Ciencias Exactas, Universidad Andres Bello, Fernandez Concha 700, Santiago, Chile}
\affiliation{Millennium Institute for Subatomic Physics at High-Energy Frontier (SAPHIR), Fernandez Concha 700, Santiago, Chile}
\author{A.~Gardikiotis\orcidlink{0000-0002-4435-2695}}
\affiliation{Physics Department, University of Patras, 265 04 Patras, Greece}
\author{S.~V.~Gertsenberger\orcidlink{0009-0006-1640-9443}}
\affiliation{Authors affiliated with an international laboratory covered by a cooperation agreement with CERN}
\author{S.~Girod}
\affiliation{CERN, European Organization for Nuclear Research, CH-1211 Geneva, Switzerland}
\author{S.~N.~Gninenko\orcidlink{0000-0001-6495-7619}}
\thanks{Corresponding author}\email{sergei.gninenko@cern.ch}
\affiliation{Authors affiliated with an institute covered by a cooperation agreement with CERN}
\author{M.~H\"osgen}
\affiliation{Universit\"at Bonn, Helmholtz-Institut f\"ur Strahlen-und Kernphysik, 53115 Bonn, Germany}
\author{V.~A.~Kachanov\orcidlink{0000-0002-3062-010X}}
\affiliation{Authors affiliated with an institute covered by a cooperation agreement with CERN}
\author{Y.~Kambar\orcidlink{0009-0000-9185-2353}}
\affiliation{Authors affiliated with an international laboratory covered by a cooperation agreement with CERN}
\author{A.~E.~Karneyeu\orcidlink{0000-0001-9983-1004}}
\affiliation{Authors affiliated with an institute covered by a cooperation agreement with CERN}
\author{E.~A.~Kasianova}
\affiliation{Authors affiliated with an international laboratory covered by a cooperation agreement with CERN}
\author{G.~D.~Kekelidze\orcidlink{0000-0002-5393-9199}}
\affiliation{Authors affiliated with an international laboratory covered by a cooperation agreement with CERN}
\author{B.~Ketzer\orcidlink{0000-0002-3493-3891}}
\affiliation{Universit\"at Bonn, Helmholtz-Institut f\"ur Strahlen-und Kernphysik, 53115 Bonn, Germany}
\author{D.~V.~Kirpichnikov\orcidlink{0000-0002-7177-077X}}
\affiliation{Authors affiliated with an institute covered by a cooperation agreement with CERN}
\author{M.~M.~Kirsanov\orcidlink{0000-0002-8879-6538}}
\affiliation{Authors affiliated with an institute covered by a cooperation agreement with CERN}
\author{V.~N.~Kolosov}
\affiliation{Authors affiliated with an institute covered by a cooperation agreement with CERN}
\author{V.~A.~Kramarenko\orcidlink{0000-0002-8625-5586}}
\affiliation{Authors affiliated with an institute covered by a cooperation agreement with CERN}
\affiliation{Authors affiliated with an international laboratory covered by a cooperation agreement with CERN}
\author{L.~V.~Kravchuk\orcidlink{0000-0001-8631-4200}}
\affiliation{Authors affiliated with an institute covered by a cooperation agreement with CERN}
\author{N.~V.~Krasnikov\orcidlink{0000-0002-8717-6492}}
\affiliation{Authors affiliated with an institute covered by a cooperation agreement with CERN}
\affiliation{Authors affiliated with an international laboratory covered by a cooperation agreement with CERN}
\author{S.~V.~Kuleshov\orcidlink{0000-0002-3065-326X}}
\affiliation{Center for Theoretical and Experimental Particle Physics, Facultad de Ciencias Exactas, Universidad Andres Bello, Fernandez Concha 700, Santiago, Chile}
\affiliation{Millennium Institute for Subatomic Physics at High-Energy Frontier (SAPHIR), Fernandez Concha 700, Santiago, Chile}
\author{V.~E.~Lyubovitskij\orcidlink{0000-0001-7467-572X}}
\affiliation{Authors affiliated with an institute covered by a cooperation agreement with CERN}
\affiliation{Universidad T\'ecnica Federico Santa Mar\'ia and CCTVal, 2390123 Valpara\'iso, Chile}
\affiliation{Millennium Institute for Subatomic Physics at High-Energy Frontier (SAPHIR), Fernandez Concha 700, Santiago, Chile}
\author{V.~Lysan\orcidlink{0009-0004-1795-1651}}
\affiliation{Authors affiliated with an international laboratory covered by a cooperation agreement with CERN}
\author{A.~Marini\orcidlink{0000-0002-6778-2161}}
\affiliation{INFN, Sezione di Genova, 16147 Genova, Italia}
\author{L.~Marsicano\orcidlink{0000-0002-8931-7498}}
\affiliation{INFN, Sezione di Genova, 16147 Genova, Italia}
\author{V.~A.~Matveev\orcidlink{0000-0002-2745-5908}}
\affiliation{Authors affiliated with an international laboratory covered by a cooperation agreement with CERN}
\author{R.~Mena~Fredes}
\affiliation{Millennium Institute for Subatomic Physics at High-Energy Frontier (SAPHIR), Fernandez Concha 700, Santiago, Chile}
\affiliation{Universidad T\'ecnica Federico Santa Mar\'ia and CCTVal, 2390123 Valpara\'iso, Chile}
\author{R.~ G.~Mena~Yanssen}
\affiliation{Millennium Institute for Subatomic Physics at High-Energy Frontier (SAPHIR), Fernandez Concha 700, Santiago, Chile}
\affiliation{Universidad T\'ecnica Federico Santa Mar\'ia and CCTVal, 2390123 Valpara\'iso, Chile}
\author{L.~Molina Bueno\orcidlink{0000-0001-9720-9764}}
\affiliation{Instituto de Fisica Corpuscular (CSIC/UV), Carrer del Catedratic Jose Beltran Martinez, 2, 46980 Paterna, Valencia, Spain}
\author{M.~Mongillo\orcidlink{0009-0000-7331-4076}}
\affiliation{ETH Z\"urich, Institute for Particle Physics and Astrophysics, CH-8093 Z\"urich, Switzerland}
\author{D.~V.~Peshekhonov\orcidlink{0009-0008-9018-5884}}
\affiliation{Authors affiliated with an international laboratory covered by a cooperation agreement with CERN}
\author{V.~A.~Polyakov\orcidlink{0000-0001-5989-0990}}
\affiliation{Authors affiliated with an institute covered by a cooperation agreement with CERN}
\author{B.~Radics\orcidlink{0000-0002-8978-1725}}
\affiliation{York University, Toronto, Canada}
\author{K.~M.~Salamatin\orcidlink{0000-0001-6287-8685}}
\affiliation{Authors affiliated with an international laboratory covered by a cooperation agreement with CERN}
\author{V.~D.~Samoylenko}
\affiliation{Authors affiliated with an institute covered by a cooperation agreement with CERN}
\author{H.~Sieber\orcidlink{0000-0003-1476-4258}}
\affiliation{ETH Z\"urich, Institute for Particle Physics and Astrophysics, CH-8093 Z\"urich, Switzerland}
\author{D.~A.~Shchukin\orcidlink{0009-0007-5508-3615}}
\affiliation{Authors affiliated with an institute covered by a cooperation agreement with CERN}
\author{O.~Soto}
\affiliation{Departamento de Fisica, Facultad de Ciencias, Universidad de La Serena, Avenida Cisternas 1200, La Serena, Chile}
\affiliation{Millennium Institute for Subatomic Physics at High-Energy Frontier (SAPHIR), Fernandez Concha 700, Santiago, Chile}
\author{V.~O.~Tikhomirov\orcidlink{0000-0002-9634-0581}}
\affiliation{Authors affiliated with an institute covered by a cooperation agreement with CERN}
\author{I.~V.~Tlisova\orcidlink{0000-0003-1552-2015}}
\affiliation{Authors affiliated with an institute covered by a cooperation agreement with CERN}
\author{A.~N.~Toropin\orcidlink{0000-0002-2106-4041}}
\affiliation{Authors affiliated with an institute covered by a cooperation agreement with CERN}
\author{M.~Tuzi\orcidlink{0009-0000-6276-1401}}
\affiliation{Instituto de Fisica Corpuscular (CSIC/UV), Carrer del Catedratic Jose Beltran Martinez, 2, 46980 Paterna, Valencia, Spain}
\author{B.~I.~Vasilishin}
\affiliation{Authors affiliated with an institute covered by a cooperation agreement with CERN}
\author{P.~V.~Volkov\orcidlink{0000-0002-7668-3691}}
\affiliation{Authors affiliated with an international laboratory covered by a cooperation agreement with CERN}
\author{V.~Yu.~Volkov\orcidlink{0009-0005-3500-5121}}
\affiliation{Authors affiliated with an institute covered by a cooperation agreement with CERN}
\author{I.~V.~Voronchikhin\orcidlink{0000-0003-3037-636X}}
\affiliation{Authors affiliated with an institute covered by a cooperation agreement with CERN}
\author{J.~Zamora-Sa\'a\orcidlink{0000-0002-5030-7516}}
\affiliation{Center for Theoretical and Experimental Particle Physics, Facultad de Ciencias Exactas, Universidad Andres Bello, Fernandez Concha 700, Santiago, Chile}
\affiliation{Millennium Institute for Subatomic Physics at High-Energy Frontier (SAPHIR), Fernandez Concha 700, Santiago, Chile}
\author{A.~S.~Zhevlakov\orcidlink{0000-0002-7775-5917}}
\affiliation{Authors affiliated with an international laboratory covered by a cooperation agreement with CERN}
\collaboration{The NA64 Collaboration}\noaffiliation
\vskip 0.25cm

\date{\today}


\begin{abstract}
Thermal dark matter models with particle $\chi$ masses below the electroweak scale can provide an explanation for the observed relic dark matter density. This would imply the existence of a new feeble interaction between the dark and ordinary matter. 
  We report on a new search for the sub-GeV  $\chi$  production through the interaction mediated by a new vector boson, called the dark photon $A'$,  in  collisions of 100 GeV electrons with the active target of the  NA64 experiment at the CERN SPS. With  $9.37\times10^{11}$  electrons on target collected during 2016-2022 runs
  NA64 probes  for the first time the well-motivated  region of parameter space of  benchmark thermal scalar and fermionic dark matter models.
   No evidence for dark matter production has been found. This  allows us to set the most sensitive limits on the $A'$ couplings to photons for masses  $m_{A'} \lesssim 0.35$ GeV, and to exclude scalar and Majorana dark matter with  the $\chi-A'$ coupling  $\alpha_D \leq 0.1$ for masses 
$0.001 \lesssim  m_\chi \lesssim 0.1$ GeV and $3m_\chi \leq m_{A'}$.
\end{abstract}


\maketitle

Thermal light dark matter (LDM) with DM particles ($\chi$) masses below the electroweak scale, $m_\chi \ll 100$ GeV,  
 is one of the most popular candidates to explain the origin of DM. In this model, dark and ordinary matter were initially in thermal equilibrium and annihilate to each other
at equal rates. However, as the Universe was expanding and cooling down, the annihilation rates fell out of equilibrium and the DM number density froze out at 
the value when equilibrium was lost. In this scenario,
the existence of a new interaction between the $\chi$ and the Standard Model (SM) is required to accomodate the relic DM density. It is remarkable that the strength of this
 interaction is such, that it allows one to search for its manifestations at present accelerators \cite{Kolb,Rubakov:2017xzr}.
 
\par In one of the most interesting cases, this interaction  could be   transmitted by a new vector boson, called dark photon ($A'$ ). The $A'$  could  have a mass in the sub-GeV mass range,  and couple to the SM via kinetic mixing  with the SM photon field   
described by the  term  $\frac{\epsilon}{2}F'_{\mu\nu}F^{\mu\nu}$, here $F'_{\mu\nu} = \partial_\mu A'_\nu-\partial_\nu A'_\mu$, where $\epsilon$ 
is  the mixing strength ~\cite{Fayet,prv,ArkaniHamed:2008qn,jr}. 
 \begin{figure*}[tbh!!]
\includegraphics[width=.8\textwidth]{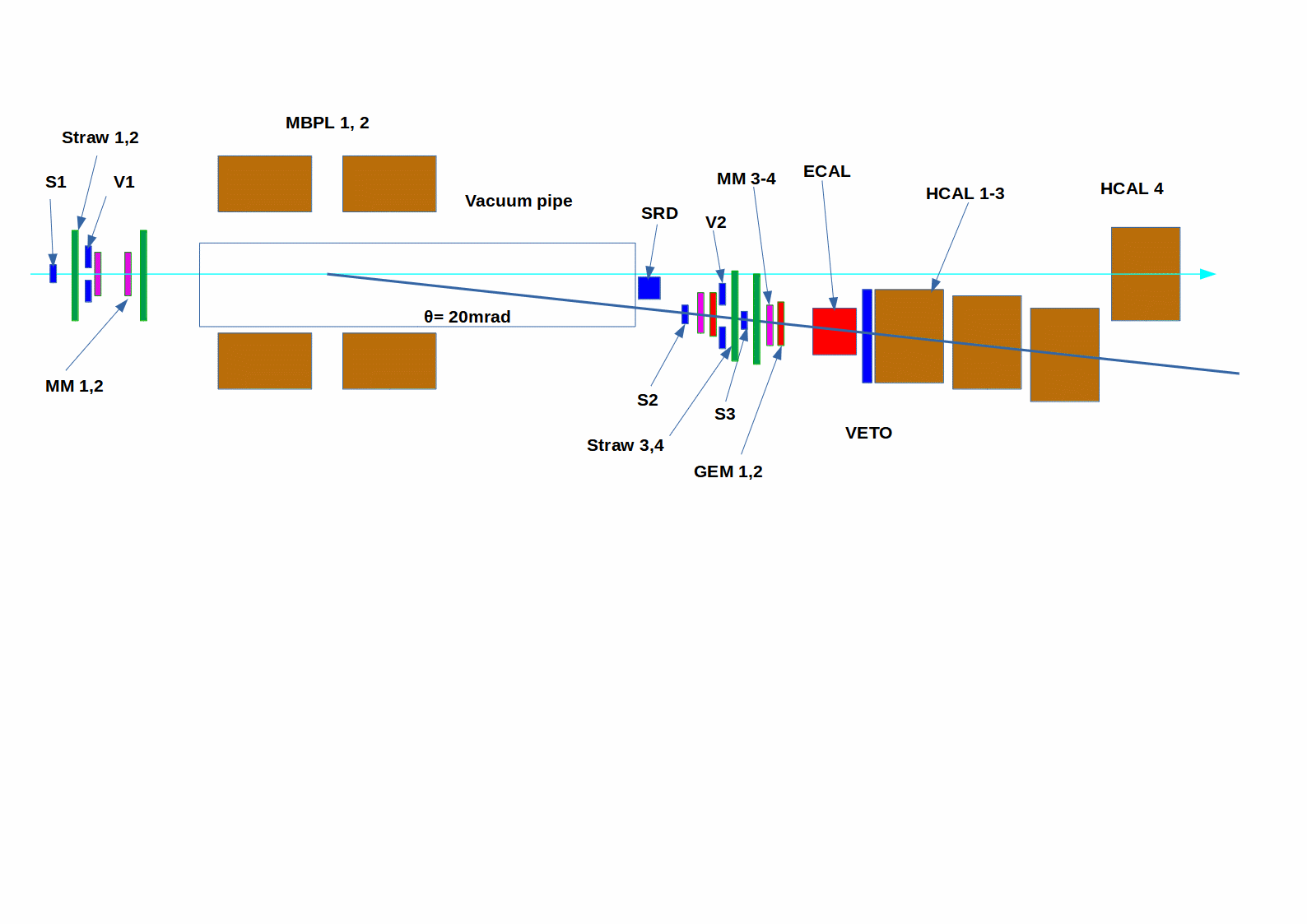}%
\vskip-4.cm{\caption{Schematic illustration of the setup to search for $\ainv$  decays of the bremsstrahlung $A'$s 
produced in the reaction  $eZ\rightarrow eZ A'$ of 100 GeV $e^-$ incident  on the active ECAL target in 2021-2022 runs.\label{setup}}}
\end{figure*} 
 The  massive $A'_\mu$  field,  associated with the  spontaneously broken  $U_D(1)$ gauge group, has  the dark  coupling strength  $e_D$ of the $U_D(1)$ gauge interactions to
$\chi$'s, which is given by  $\mathcal{L}_{int}=  - e_D A'_\mu J^\mu_D$, where $J_D$ is the dark matter current. 
The mixing term  results in the interaction $\mathcal{L}_{int}= \epsilon e A'_{\mu} J^\mu_{em}$ 
of the $A'$  with the electromagnetic (e-m) current $J^\mu_{em}$ with a strength $\epsilon e$, 
where $e$ is the e-m coupling and $\epsilon \ll 1$ \cite{Okun:1982xi,Galison:1983pa,Holdom:1985ag}. Small values of $\epsilon$ 
can be obtained in grand unified theories  from loop effects  of particles having both the dark $U_D(1)$ and SM  $U(1)$  charges with  a typical one-loop  value  $\epsilon = e e_D/16\pi^2 \simeq 10^{-2}-10^{-4}$ or from  two-loop contributions resulting  in 
$\epsilon \simeq 10^{-3}-10^{-5}$ \cite{Holdom:1985ag}. 
\par 
If the $A'$  is the lightest state in the dark sector,  then it would decay mostly to a visible state, i.e. to SM leptons or hadrons \cite{pdg, apex, merkel, babar1, phenix, na48, kloe3, na64prl18, na64prd20}. However, if the decay $A'\to \chi \chi$ is kinematically allowed,  the $A'$  would dominantly  decay invisibly  into $\chi$'s provided 
 $m_\chi<\ma/2$ and  $e_D >\epsilon e$.  Various LDM  models motivate the existence of sub-GeV $\chi$'s which could be either scalar, Majorana,  or  pseudo-Dirac  particles coupled to the $A'$ 
\cite{Fayet:2007ua, Pospelov:2008zw,deNiverville:2011it,Izaguirre:2014bca,Iza2015,Iza2017,pat}.  Models introducing the invisible  $A'$, i.e. that invisible decay mode is predominant,  $\Gamma (A'\rightarrow \bar{\chi}\chi)/\Gamma_{tot} \simeq 1$,  are subject to various experimental probes leaving,  however,  a  large parameter area still to be explored. 
\par Imposing the thermal freeze-out condition of DM annihilation into visible sector through $\g-A'$  mixing allows one to predict values of  the parameter 
\begin{equation}
y= \alpha_D \epsilon^2 \Bigl( \frac{m_\chi}{\ma}\Bigr)^4
\label{eq:y}
\end{equation}
which defines the annihilation cross section and  hence the relic DM density, and also relates the dark coupling  $\alpha_D = e_D^2/4\pi$  and mixing $\epsilon$ by 
\begin{equation}
\alpha_D \simeq 0.02 f \Bigl( \frac{10^{-3}}{\epsilon}\Bigr)^2\Bigl( \frac{\ma}{100~ {\rm MeV}}\Bigr)^4
\Bigl( \frac{10~{\rm MeV}}{m_\chi}\Bigr)^2
\label{eq:alphad}
\end{equation}
where  the parameter $f$ depends on $m_{A^`}$ and $m_{\chi}$ \cite{Kolb}.
For $\frac{m_{A^`}}{m_{\chi}} =3$,  $f \lesssim 10$ for a scalar \cite{deNiverville:2011it}, and $f\lesssim 1$ for a  fermion case \cite{Izaguirre:2014bca}.
The accessibility of the predicted $y$ and $\alpha_D$  values in direct searches and at accelerator experiments provide an important complementarity between different techniques\cite{comp}. Thus, motivating a worldwide effort towards dark forces and other portals between the visible and dark sectors; see, e.g. Refs. \cite{Essig:2013lka,report1,report2, pbc-bsm, pbc, fips, snowm, bb, gori, gl1, gl2, berlin, Lee:2014tba,deNiverville:2011it,Diamond:2013oda,hd,Essig:2013vha, Batell:2009di, e137th, na64prl, minib2018, na64prd,na64prl19,babarg-2,na62}. 
In particular, the predicted ($\epsilon, ~y, ~\ma,~\mc$) parameter space which can be probed at the beam energies of the CERN SPS. 
 
\par In this Letter, we report new results on the search for the invisible $A'$ mediator and light-dark matter in the fixed-target experiment NA64 at the CERN SPS \cite{ufn,pc}, 
obtained from the combined statistics of 2016-2022 runs.
The method we chose for the search was proposed in Refs.~\cite{Gninenko:2013rka,Andreas:2013lya}. It is 
  based on the detection of the missing energy, arising from prompt decays $A'\to \chi \chi$ of the 
hard bremsstrahlung $A'$   produced in the process $e^-Z \to e^- Z A'$  of high-energy electrons scattering in the active beam dump target.  Another $A'$  production mechanism considered in this work is through the resonant annihilation of secondary positrons from the e-m shower developed in the target 
with its atomic electrons, $\ee \to A'\to \chi \chi$ \cite{luca}. Due to the resonant cross-section enhancement, it provides a strong increase to the signal yield, and thus to the sensitivity of the search, in particular for the high-mass region, $\ma \gtrsim m_\mu$ \cite{na64res}. 
The advantage of the NA64 approach compared to the classical beam dump ones  \cite{deNiverville:2011it,Izaguirre:2014bca,Diamond:2013oda,Batell:2009di, Gninenko:2012prd,Gninenko:2012plb}  is that its sensitivity is proportional to $\epsilon^2$,  associated with the $A'$ production and its subsequent prompt invisible decay, while in the latter case, it is proportional to $\epsilon^4\alpha_D$, with $\epsilon^2$ associated with the $A'$ production in the dump and  $\epsilon^2 \alpha_D$ coming from the $\chi$ particle scattering in the far detector \cite{Gninenko:2013rka}.
\par  The  NA64 detector upgraded for a more sensitive LDM search in 2021-2022 runs is   schematically shown in Fig.~\ref{setup}. It  employs  the optimized H4  100 GeV  electron beam at CERN SPS,  
which  has a maximal intensity $\simeq 10^7$ electrons per  SPS spill of 4.8 s \cite{h4}.  The beam is defined by the scintillator (Sc)  counters $S_{1-3}$ and a veto counter $V_{1}$. 
A  magnetic spectrometer is used to reconstruct the momentum of the incoming  $e^-$'s with the precision $\delta p/p \simeq 1\%$ \cite{Banerjee:2015eno}. The spectrometer consists  of two  consecutive dipole magnets MBPL$_{1,2}$ with the total magnetic field of $\simeq$7 T$\cdot$m  and a  low-material-budget tracker composed of a set of  two upstream Micromegas 
(MM$_{1,2}$)  and two straw-tube chambers (ST$_{1,2}$) and two downstream MM$_{3-4}$, ST$_{3,4}$, and GEM$_{1,2}$ stations. 
The synchrotron radiation (SR) emitted in the MBPL magnetic field and detected with an SR detector (SRD) was used for electron identification. The SRD is an array of a  PbSc sandwich calorimeter of a fine segmentation \cite{Gninenko:2013rka, na64srd}. With this technique the initial admixture of the hadron contamination in the beam 
$\pi/e^- \lesssim 10^{-2}$ was further suppressed to $\simeq 2\times 10^{-5}$ \cite{na64beam}. 
 Downstream the setup  was  equipped with an active dump target, an  e-m calorimeter (ECAL), 
 for  measurement of the recoil electron energy  $E_{ECAL} $ and the transverse and longitudinal shape of the corresponding e-m shower. The ECAL  
 was  a  matrix of $5\times 6 $  Shashlik-type modules assembled from  Pb and Sc plates  of $\simeq 40$ radiation  lengths ($X_0$),  with the first 4$X_0$ serving as a preshower detector.   Finally,  the ECAL followed by  a  high-efficiency veto counter VETO, and a massive, hermetic hadronic calorimeter (HCAL) of  three modules HCAL$_{1-3}$,  $\simeq 30$ nuclear interaction lengths in total to  veto muons or hadronic secondaries produced in the $e^-$ nuclei interactions  in the target.  Zero-degree HCAL$_4$ was used to reject 
 beam electrons accompanied by neural secondaries. 
\begin{figure}[tbh!!]
\hspace{-0.cm}{\includegraphics[width=0.45\textwidth]{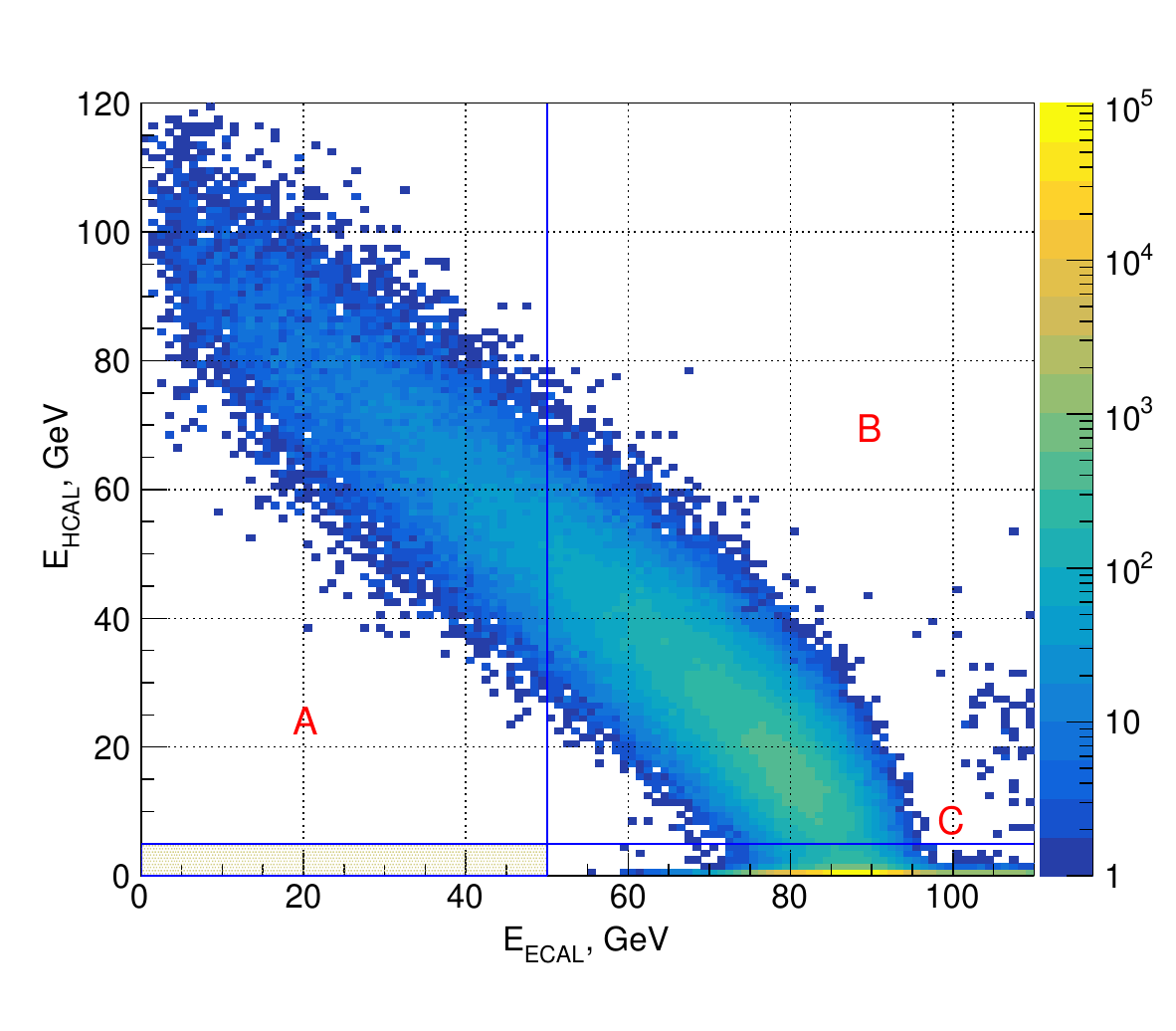}}
\caption{
The measured  distribution of events in the ($E_{ECAL}$;$E_{HCAL}$) plane  after applying all selection criteria. The shaded area is the signal box, with the size along the $E_{HCAL}$ axis increased for illustration purposes. The side bands $A$ and $C$ are the ones  used for the background estimate inside the signal region.  }
\label{ecvshc}
\end{figure}   
\par Our data were collected in several runs, during two periods with the trigger  requiring  the ECAL energy  $E_{ECAL} \lesssim 90$ GeV. 
The first period had $2.83\times 10^{11}$ electrons on target (EOT) accumulated during 2016-2018 runs ( (hereafter called respectively  runs I-III) \cite{na64prd,na64prl19}. The second, with  
2021 run (run IV) \cite{na64prl22} and 2022 run (run V),  had  $6.54\times 10^{11}$ EOT collected with the beam intensity in the range  $\simeq(5-7)\times 10^6$   e$^-$ per spill.  Data with a total of $9.37\times 10^{11}$ EOT from these five runs were processed with  selection criteria and  combined as described below. 
\par  A  Geant4 \cite{Agostinelli:2002hh, geant} based Monte Carlo (MC) simulation package DMG4 \cite{dmg4} is used to study the performance of the detector,  signal acceptance, and  background level, as well as the analysis procedure  including  selection of  cuts and sensitivity estimate. 
\begin{figure}[tbh]
\begin{center}
\includegraphics[width=0.45\textwidth]{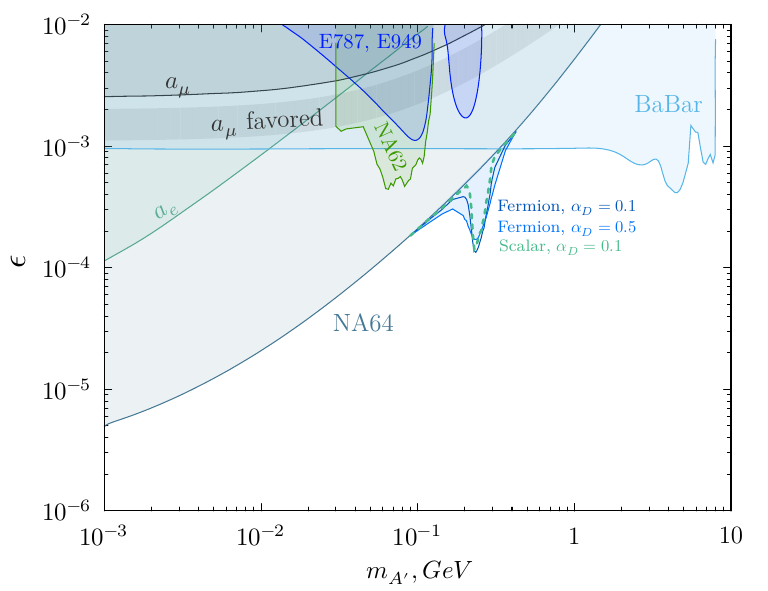}
\caption {The NA64 90\% C.L. exclusion region in the ($m_{A'}, \epsilon$) 
plane.   Constraints from the  
 E787 and  E949  \cite{hd,Essig:2013vha}, $BABAR$ \cite{babarg-2} and NA62 \cite{na62} experiments, 
 from the consideration of the anomalous magnetic moment of electron  $\alpha_e$ \cite{han, bou, mor, aoy},   as well as the favored area  explaining the  $\alpha_\mu$ anomaly with the $A'$ contribution  \cite{pdg} are also shown. 
 For more limits from indirect searches and 
 planned measurements; see,  e.g.,  Refs. \cite{report1,report2, pbc-bsm}.
  \label{exclinv}}
\end{center}
\end{figure} 
  To maximize the signal acceptance and to minimize  background, the following   selection criteria  were used:
(i) The incoming track  should have the momentum $100\pm 10$ GeV.
(ii) The track angle with respect to  the deflected beam axis should be within 3 mrad  to reject large angle events from the upstream $e^-$ interactions. 
(iii)  The detected SR energy should be within the range $\simeq 1-100$ MeV  emitted by $e^-$s and in time with the trigger. 
(iv)  The  longitudinal and lateral shape of the ECAL cluster should be  consistent with the one expected for the signal event \cite{gkkk}.
(v) There should be no multiple hits in the ST$_{3,4}$ chambers and  no activity in VETO. This was an effective  cut against the electroproduction of charged secondaries  in the upstream beam material. The measured  distribution of $\simeq 9.6 \times 10^5 $ events in the ($E_{ECAL}$;$E_{HCAL}$) plane that passed  these criteria from combined runs IV and V is shown in Fig.~\ref{ecvshc}.  The candidate events were requested  to have the missing energy  $E_{miss}  \gtrsim50 $ GeV. The optimal signal box 
was defined based on the energy spectrum calculations for $A'$s emitted by $e^\pm$ from the  e-m shower generated by the primary  $e^-$s in the  target \cite{gkkk, gkkketl} and the expected background level for each particular run. Events originated from the rare QED dimuon production in the target were  used as a benchmark reaction  allowing us  to verify the reliability of the MC simulation, correct  the signal acceptance vs the $A'$ energy,  cross-check  systematic uncertainties  and background estimate \cite{na64prd,na64prl19}. 
The $A'$ acceptance was evaluated from simulations and  $e^-$ data sample taking into account the selection efficiency for the longitudinal and transverse 
  e-m shower shape in the target from signal events \cite{gkkk}. The energy corrections were extracted using the measured spectra of recoil $e^-$ in the ECAL from 
  dimuon events. The $A'$ production cross section for the bremsstrahlung reaction was obtained with exact tree-level calculations as described in
   Ref. \cite{gkkketl} with 
   uncertainty  $\simeq 10\%$ \cite{na64prd,na64prl19}, while the $A'$ yield from the resonant process was calculated as described in Ref. \cite{na64res}.
\begin{figure*}[tbh!]
\hspace{-0.5cm}{\includegraphics[width=.44\textwidth]{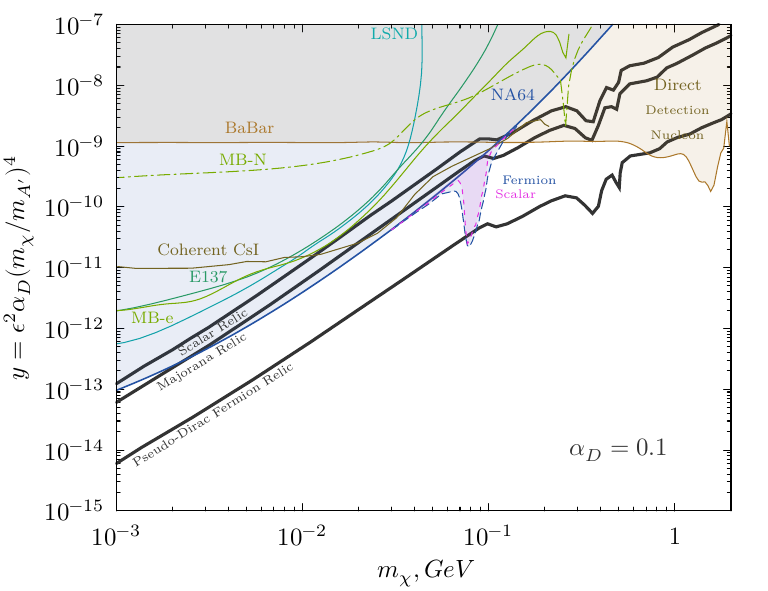}}
\includegraphics[width=.44\textwidth]{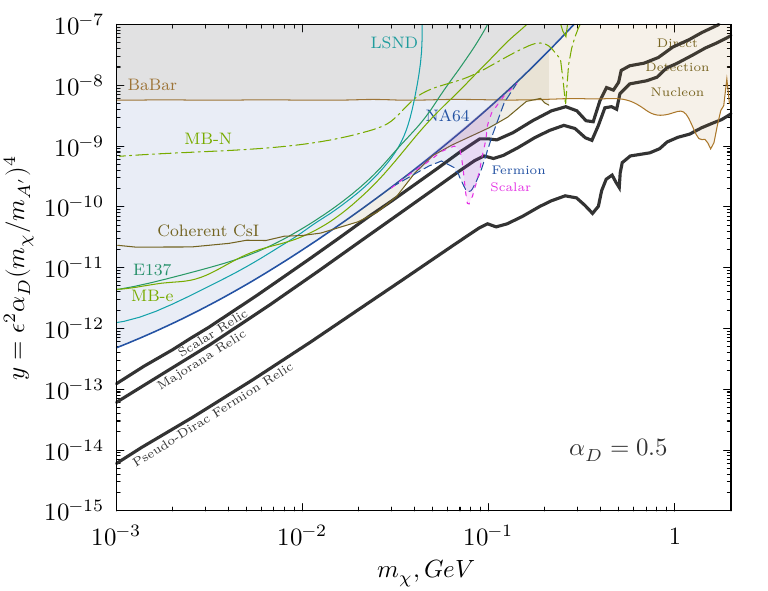}
\includegraphics[width=0.47\textwidth]{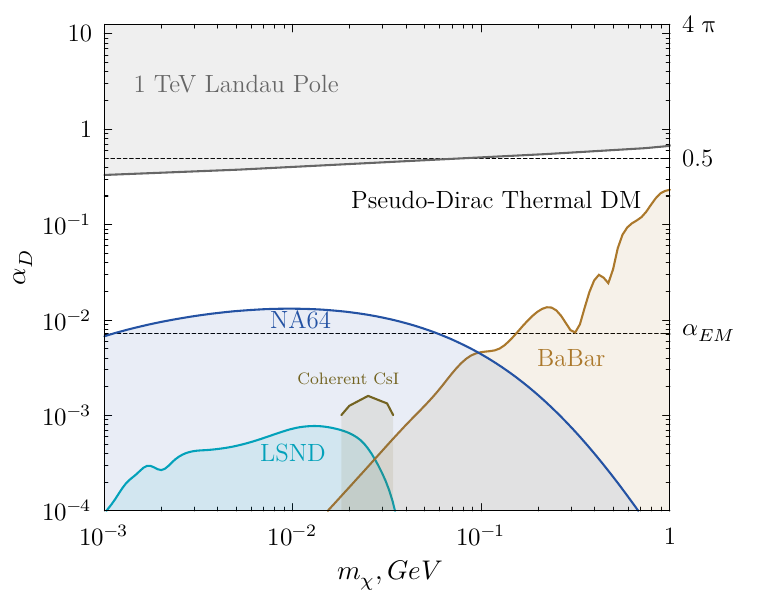}
\includegraphics[width=0.47\textwidth]{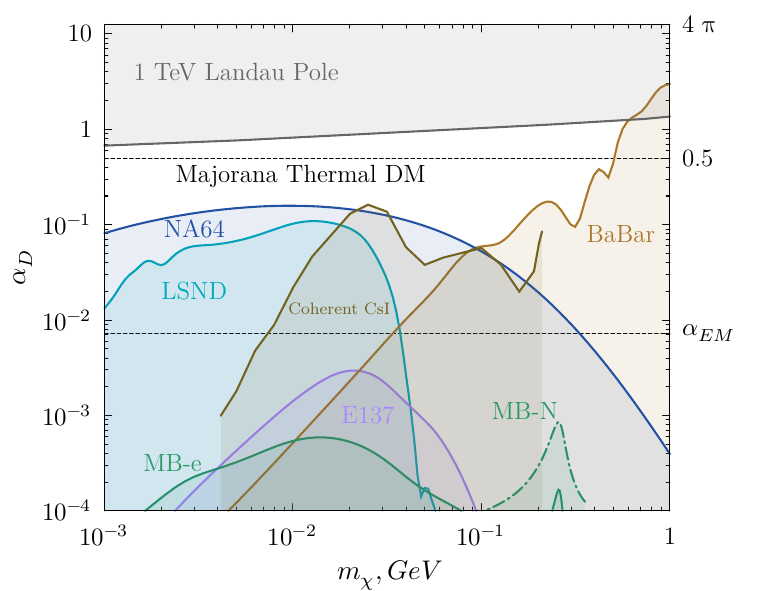}
\caption{The top row shows the NA64 limits  in  the (y;$m_{\chi}$) plane obtained for $\alpha_D=0.5$ (left panel) and $\alpha_D=0.1$ (right panel)
assuming   $m_{A'}=3 m_{\chi}$, from the full 2016-2022 data set.
The bottom row shows the NA64 constraints in the ($\alpha_D$;$m_{\chi}$) plane  on  the pseudo-Dirac (left panel) 
and Majorana (right panel) DM. The limits are shown   
 in comparison with bounds obtained in Refs.\cite{report1, report2, Izaguirre:2014bca,Iza2015,Iza2017}  from  the results of the 
LSND~\cite{deNiverville:2011it,Batell:2009di},   E137 \cite{e137th}, MiniBooNE \cite{minib2018}, $BABAR$ \cite{babarg-2}, COHERENT \cite{coh},  and direct detection \cite{mardon} experiments. The favored parameters to account for the observed relic DM density for the  scalar, pseudo-Dirac  and Majorana  type of light DM are shown as the lowest solid line
in top plots; see, e.g. \cite{berlin}.}
\label{yvsm}
\end{figure*}   
\begin{table}[tbh!] 
\begin{center}
\caption{Expected background  for  2021-2022 runs}\label{tab:bckg}
\vspace{0.15cm}
\begin{tabular}{lr}
\hline
\hline
Background source& Background, $n_b$\\
\hline
(i) dimuons losses or decays in the target &$ 0.04\pm 0.01$\\
(ii)  $\mu,\pi,K\to e +....$ decays in the beam line & $ 0.3\pm 0.05$ \\
(iii) lost $\g, n, K^0$  from upstream interactions  &$0.16\pm 0.12$\\
(iv) Punch-through leading $n,K^0_L $ & $<0.01$\\
\hline 
Total $n_b$ (conservatively)   &    $0.51\pm 0.13$\\
\hline
\hline 
\end{tabular}
\end{center}
\end{table}
\par Several  processes  shown in Table \ref{tab:bckg} contribute to background: (i) dimuons losses due 
to their inefficient detection  or decays in the target. It was estimated from the measurements of the single muon detection efficiency and the number of observed dimuons;  (ii) decays in flight of mistagged  
$\mu, \pi$, $K$. It was evaluated from the simulations and measurements of the beam composition \cite{na64beam};  (iii) escaping large-angle neutrals from the electroproduction in the beam line due to the insufficient HCAL coverage. Compared to the first period of data taking,  this background 
was significantly reduced in 2021-2022 runs by increasing the HCAL acceptance by moving it $\simeq 3$ m  upstream.  Additionally, to minimize interactions in the beam line the amount of dead material was further reduced;  the last background source is  (iv) punch-through of leading neutral hadrons $(n, K^0_L)$ from the $e^-$  interactions in the target.  It was evaluated from the direct measurements of punch-through events \cite{na64alp}.
After applying the selection cuts, we expected mostly background events of type (iii)  to remain in the data. Their number was evaluated  from the data itself by the extrapolation of events  from the sideband 
 $C$ ($E_{ECAL} > 50~{\rm GeV }; E_{HCAL} < 1~{\rm GeV }$) shown in  Fig. \ref{ecvshc} into the signal region and assessing the systematic errors by varying the fit functions. The shape of the extrapolation functions was taken from the analysis of the data and cross-checked with simulations of the $e^-$  hadronic interactions in the beam. Finally, the estimated background inside the signal region was  0.51$\pm$0.13 events.  Compared to the 2016-2018 runs \cite{na64prl19},  background was further rejected by a factor $\simeq 6$.  After determining all the selection criteria and background levels, no event is found in the signal region. 
\par To obtain  upper limits on the mixing strength,  runs I-V were analyzed simultaneously using the technique of multi-bin limit setting
 based on the RooStats package \cite{root}. First, the background levels, efficiencies, their corrections and uncertainties were used to optimize the ECAL energy cut for the signal box,  by comparing sensitivities,  defined as an average expected bound calculated using the profile likelihood ratio method for each run. 
 For this procedure, the expected number of background events for each run,  obtained from the data extrapolation to the signal region with errors estimated from the variation of 
the extrapolation functions,  were the most important inputs. For the data samples from the 2016-2021 runs we used  the previously optimised value of the 
ECAL energy cut of 50 GeV \cite{na64prd, na64prl19,na64prl22}.   For the long 2022 run, the optimal cut was selected in the range  $E_{ECAL} \lesssim 47-50$ GeV, slightly depending on the run conditions and detector performance during the data taking.   
 The overall signal efficiency, given by the product of efficiencies accounting for the geometrical acceptance, the  track, SRD, VETO  and HCAL signal reconstruction, 
was  in the range 0.4-0.5. The VETO (0.94) and HCAL(0.95)  efficiencies  were defined as a fraction of events below the zero-energy thresholds, with the loss mostly due to pileup in high-intensity runs.
\par  The combined 90\%  confidence level (C.L.)  upper limits for  mixing strength $\epsilon$ is obtained 
 by applying the modified frequentist approach for confidence levels, considering the profile
likelihood ratio as a test statistic in the asymptotic approximation \cite{junk,limit,Read:2002hq}. The number of events in the signal box is the sum of expected events from all five runs:
\begin{equation}
\Na = \sum_{i=1}^{5} N_{A'}^i = \sum_{i=1}^{5} n_{EOT}^i  \epsilon_{A'}^i n_{A'}^i(\epsilon,\ma, \Delta E_{e})
\label{nev}
\end{equation}
where $\epsilon_{A'}^i$ is the signal efficiency in run $i$, and $n_{A'}^i(\epsilon,\ma, \Delta E_{e})$ is the number of
signal events per EOT produced in the energy interval $\Delta E_{e} $. 
Signal events for each $i$th entry in Eq.(\ref{nev}) are simulated and reconstructed with the same selection criteria and efficiency corrections as for the data sample from run $i$.
The combined 90\% C.L. exclusion limits on $\epsilon$ as a function of the $A'$ mass,  calculated by taking into account the estimated backgrounds and systematic errors
$\sim 15\%$ for the  $\epsilon_{A'}^i$  dominated by the $\sim 10\%$ uncertainty in the $A'$ yield  \cite{na64prd} can be seen in Fig.~\ref{exclinv}.
\par Using obtained limits,  Eqs.(\ref{eq:y}) and \eqref{eq:alphad}, one can get constraints on the LDM models, which are shown 
 in the ($y$;$m_{\chi}$) and ($\alpha_D$;$m_{\chi}$)  planes in Fig.~\ref{yvsm} for  $m_{\chi} \lesssim 1$~GeV.   
The favored $y$ parameter curves for scalar,  pseudo-Dirac (with a small splitting) and Majorana scenario  of  LDM  obtained by taking into account the observed relic DM density \cite{berlin} are also shown on the same plot.  One can see that our results are already starting to probe the  $y;m_\chi$  parameter space predicted for the 
benchmark values  $\alpha_D= 0.1$  and  $m_{A'}=3 m_{\chi}$ \cite{report2, pbc-bsm} providing the best limits in comparison with bounds from other experiments. Note,  that choice of $\alpha_D=0.5$ value is still compatible with the constraints  
  obtained from consideration of the running $\alpha_D$ \cite{gkkketl,davou}.   
 The limits on $\alpha_D$ for the case of pseudo-Dirac fermions shown in Fig.~\ref{yvsm} (left panel in the bottom row)  were calculated 
  by taking the value $f=0.25 $,  while for the Majorana case (right panel) the value   $f=3$  in Eq.(\ref{eq:alphad}) was used, see  Ref.\cite{na64prl19}.
\par  In summary, with the combined statistics of the 2016-2022 runs,  NA64 started probing the very exciting region of parameter space predicted by benchmark LDM scenarios. 
 Using the missing energy technique, NA64 places  stringent bounds on $\epsilon,~y, ~ \alpha_D$ 
 which are one or more orders more sensitive than the other searches in the  mass range $0.001\lesssim \ma \lesssim 0.35$ GeV \cite{pdg}. 
The scalar and Majorana models for the coupling  $\alpha_D \leq 0.1$,  the mass range  $0.001\lesssim m_\chi \lesssim 0.1$ GeV and  $3m_\chi \leq m_{A'}$ are excluded.   Further detector upgrade will enable improvements in sensitivity and coverage of the LDM  parameter space. 
 \begin{acknowledgments}
\par  We gratefully acknowledge the support of the CERN management and staff, as well as  contributions from  
HISKP, University of Bonn (Germany), ETH Zurich and SNSF Grant No. 169133, 186181, 186158, 197346 (Switzerland),  
ANID - Millennium Science Initiative Program - ICN2019 044 (Chile),  RyC-030551-I and PID2021-123955NA-100 funded by MCIN/AEI/ 10.13039/501100011033/FEDER, UE (Spain). 
This result is part of a project that has received funding from the European Research Council (ERC) under the European Union's Horizon 2020 research and innovation programme, Grant agreement No. 947715 (POKER).
\end{acknowledgments}

\end{document}